# Agent Based Trust Management Model Based on Weight Value Model for Online Auctions

E.Sathiyamoorthy, N.Ch.Sriman Narayana Iyenger & V.Ramachandran

School of Computing Sciences,
VIT University,Vellore-632014 (Tamilnadu), INDIA
sathya.probing@gmail.com, nchsniyengar48@gmail.com
&
Vice-Chancellor, Anna University , Trichy-620 024 (Tamil Nadu), , INDIA
rama@annauniv.edu

*Abstract*

*This paper is aimed at the stipulations which arise in the traditional online auctions as a result of various anomalies in the reputation and trust calculation mechanism. We try to improve the scalability and efficiency of the online auctions by providing efficient trust management methodology considering several factors into consideration. A comparison between the performance of the auctions system with and without the agent methodology is done with good results.*

## 1. Introduction

An online auction is simply defined as a virtual marketplace hosted on the Internet to match buyers and sellers of goods around the globe regardless of the physical limitations of traditional auctions such as geography, presence, time, and space. Online auctions operate different protocols including English, Dutch, First-Price Sealed Bid and Vickery with different properties for each one of these protocols. Here in this system we implement the different auctions and also help the users to decide upon the optimal price of the product considering a specific set of parameters. Our main interest would be to work on the reputation and trustworthiness of the participants in the online auctions. Since the participants in the auctions do not know the details of the seller, they have no other option rather than to trust the feedback mechanisms provided implicitly. This leaves a lot of scope for improvement in this field regarding the exact portrayal of real time scenarios of online auctions. There is also the problem of the need for constant human monitoring.

This paper is organised as follows. The Section II deals with previously achieved progress and contemporary work done in the field of e- commerce with emphasis on online auctions. In Section III, the framework of the system is dealt with. In section IV, the agent technology is explained in detail. In the fifth section, trust calculation using the proposed formula is done. In the sixth section, the results and conclusions to depict the scalability and efficiency of the system are shown.

## 2. Related work

The present e-commerce applications concerning online auctions either use the accumulative model or the ratio methodology to calculate the trust worthiness of a client. The accumulative model uses the summation of the feedbacks of other users which can be either 0 or 1 or -1. The ratio model refers to the ratio of the positive feedbacks to the total number of feedbacks. But these both neglect the possibility of malicious users and do not take the credibility of the person who is rating the other client into account. The accumulative model makes use of the summation of the reputations which are calculated over time. This is valid in case of ratio model except for the summation part.

An example of how the trustworthiness is calculated on most of the auction sites is as follows. They use a feedback system, which allows users of the online auction system to rate sellers and buyers based on the outcome of the transactions. There are three types of feedback that a user can leave. Suppose that there are $N$ number of clientele participating in the transactions on a particular site. Since the feedback values provided are given by a subset of these N clients, these values are scalars which can be any one in the set of 3 i.e. {-1, 0, 1}. So the scalars S= {1, …, N}. $F_{ij}$ denotes the feedback value provided by i for j.

$F_{ij}$ = Positive: +1 points. This denotes that the transaction went smoothly between *i* and *j* and *i* was satisfied with the services of *j*.
$F_{ij}$ = Negative: -1 points. This denotes that *i* was not happy with the outcome of the transaction with *j*.
$F_{ij}$ = Neutral: 0 points. This denotes that the transaction went through ok but it could have been better.

Once a user has reached a certain amount of points, that user receives a star indication his / her trustworthiness. There are several stars in which a user can received base on the amount of points they have earned. The colors of the stars can be yellow, blue, Turquoise, purple, red, green, and several others. This is decided based on the amount of points obtained.

Based on this system, a user can decided on whom to do business with by looking up a person's feedback. A user that has a majority of positive feedback will do more business than a user that has a majority of negative feedback. These records however, are stored on a central server that is maintained by the organization.

But, the above method does not take the weight values of rater's into account. There is also the problem of reputation squeeze which is caused because all the products are treated equally without any consideration to the transaction value. E.g. It treats the feedback given to a 20$ transaction and the feedback given to a 2000$ transaction the same way. The other problem is collusive rating where the seller's friends disguise as buyers to increase the transaction value. To avoid this problem a new method of considering weight values for the rater was found. The introduction of vectors was done. Here the weight values are in the closed interval of 0 to 1 and these are used to calculate the final feedback values which are used to compute the trust.

But, the semantics of this computed value were not clear and it was difficult to interpret it. It could be interpreted as a probability of behavior, and also could be interpreted as the trustworthiness of the user too. Thus a problem arises. The defense which is proposed against a collusive attack is very crude and depends on many random factors. When a collusive user does a good rating of the target user, the member of a collusive user is assuming that there is no similarity in the evaluation of users other than the target user. On the contrary, it is thought that no collusive attack exists when there is a similarity in the evaluation. The defense against any possible collusive attacks was left to the tolerance of the system, putting weight on the evaluation value by using this similarity.

These present mechanisms do not take the dynamic nature of these online auctions into consideration and thus several subtle features like the decay of feedback and recent trust values etc are ignored. The calculation of trust at run time can be very scalable in certain situations but it leads to congestion in the network in most of the cases due to reasons like people who bid opportunistically.

The present day auction sites mostly require the users to constantly monitor the proceedings of the auction in which they are involved. This can be a rather tiresome business. To avoid this problem we use concept of agent technology. A comparison between the performance of the auctions system with and without the agent methodology is done with agents. This comparison is based on several factors such as the time taken for the auction, the original price obtained versus the price expected etc.

## 3. Framework

The frame work consists of various components like Auction agents, Trust management agent. An agent is defined as a software entity that can perform information-related tasks without ongoing human supervision. This methodology is achieved by the usage of JADE agent technology.

The auction agent deals with the implementation of the type of auction specified. It also deals with the support provided to the user regarding the optimal price taking the parameters of quantity and priority of sales into account. The auction agent takes responsibility for the implementation of the negotiation autonomously. It takes the amount as input from the user and uses this as a threshold and bids on behalf of the user and thus the user need not monitor the proceedings at all times. The auction agent is unbiased in its functioning and strives to provide the best results to the entire clientele population. The auction agents are spawned separately for each new user arriving to participate in the auction based on their customized preferences.

The trust management agent is used to calculate the reputation of all the users taking part in the auctions either directly or indirectly. The reputation levels depict the trustworthiness of the person who is hosting the product. By trustworthiness, we mean that we are finding the weight value of the rater or user. This can be done on the basis of several factors like feedback decay, recent price, rater's trustworthiness etc. In feedback rating the rater generally rates the host according to several critical attributes which may be the quality of service provided, the type of technical support provided, the delivery of the product, the item's condition on delivery etc. Thus feedback rating as considered in the present methodologies is not a scalar but a vector quantity. Thus it can lie in the closed interval of -1 to 1 and not strictly one of the extremes.

We must also take into consideration the other factors of influence such as the trust value calculated in the nearest past and the time delay from the recent trust value, the value or the final outcome of the transaction for a better portrayal of the real time auction scenario. Thus, the trust management agent calculates the reputation and trust worthiness of the client taking into consideration the number of participants from the time t-1 to t. It is pretty obvious that if the seller has good sales as depicted by his trust value and experience value, then we can be sure that he is trustworthy. The weight values are also normalised. The merits in the ratio methodology have also been incorporated. Thus the trust management agent involves the functionalities of all the previous methods such as the accumulative, the ratio and the weight value models.

Three different types of auction protocols have been implemented here. They are the English, The Dutch and the Vickrey. The Auction agents spawn different types of agents for each type of auction protocol's modus operandi. The user interface is very friendly and no previous knowledge of the system is required. The database structure is distributed for the auction agents and it is centralized for the trust management agents for the purpose of performance enhancement.

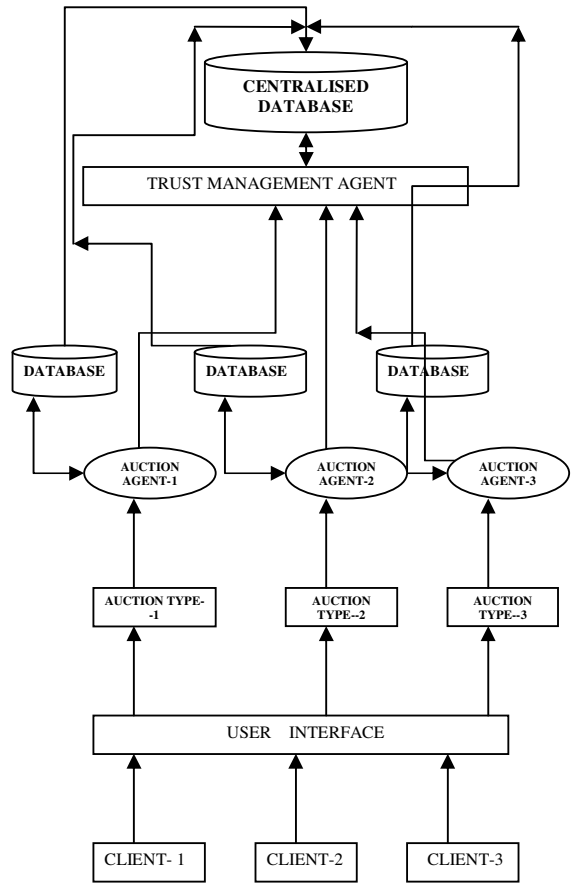

Fig.1: Frame work

When the data required for an operation is not found in the local databases then, the query is redirected to the centralised database. This reduces the burden of manipulating and working with large amounts of data when there is no particular requirement. This reduces a lot of load on the server as the database has both centralized and de centralized architectures.

Fig.2 represents the sequence diagram which depicts the entire flow of the auction system which includes the working methodologies of the various agents such as the buyer agent, seller agent, trust management agent etc. There are three methodologies depicted in the following figure. They are the purchasing part, the hosting of an auction and the final part is the feedback and trust calculation methodology. These are accomplished by the various agents which are made use. TMA represents the Trust Management agent.

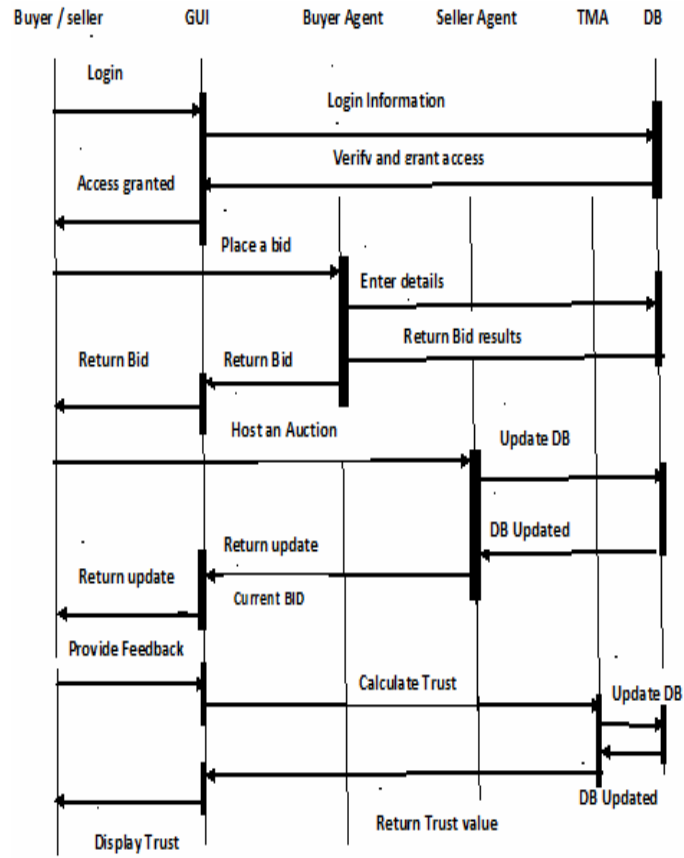

Fig.2: Sequence Diagram

Figure 3 illustrates the Buyer Agent Model. According to this model, a buyer agent contains the Bidder modules. The bidder acts based on its buyer agent's strategic model. It must decide in which e-auction to play, when to bid and how much should be the bidding value for a product. The Bidder checks the bidding status on e-auction site i.e. information such as the current bidding value, the players bidding behavior, and the closing time. An automatic bidding process should act rationally, correctly, and fast. In addition, it should behave properly when an unexpected situation occurs. The strategic behavior model continues until the buyer agent closes the deal or leaves the auction. Even though in an e-auction we have a finite number of players, we cannot estimate a priori the amount of bidders. We also consider a finite number of decisions since there is time limitation in the bidding process.

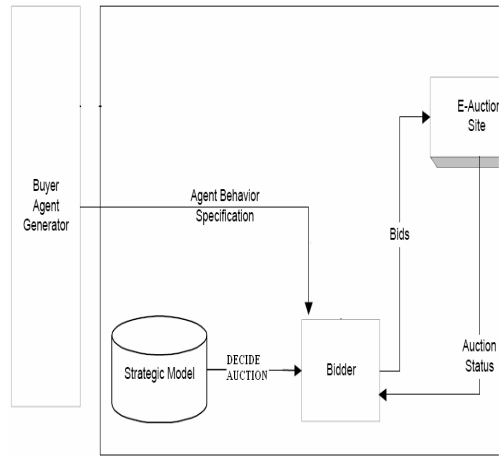

**Fig.3: Buyer Agent Model**

## 4.1 Agent Technology

Agent technology is used for the task of automating e-commerce business processes in view of bringing efficiency, scalability and profitability to businesses and individual users. Agent technologies can be used to depict the real world scenarios in the field of e-commerce onto the virtual screen.

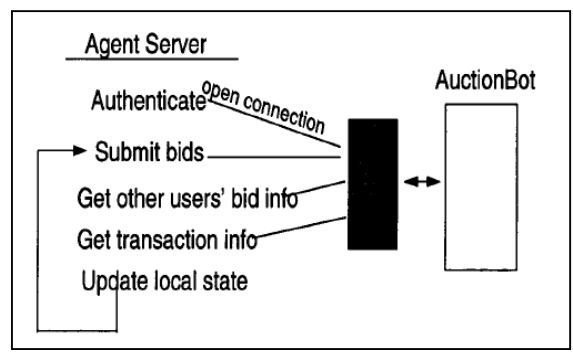

Fig.4: Communication with Agent

JADE 3.1 agent platform is used to implement the agents. JADE is one of the best modern agent environments. It is open-source and is FIPA compliant and runs on a variety of

21

operating systems which include Windows and Linux. It's very scalable because as the load on the server increases the Jade agent technology can be used for the load balancing phenomenon. The negotiations which take place between the host of the auction and the customer are performed in Jade containers. The Main container and the other containers are used to take care of the scalability issues.

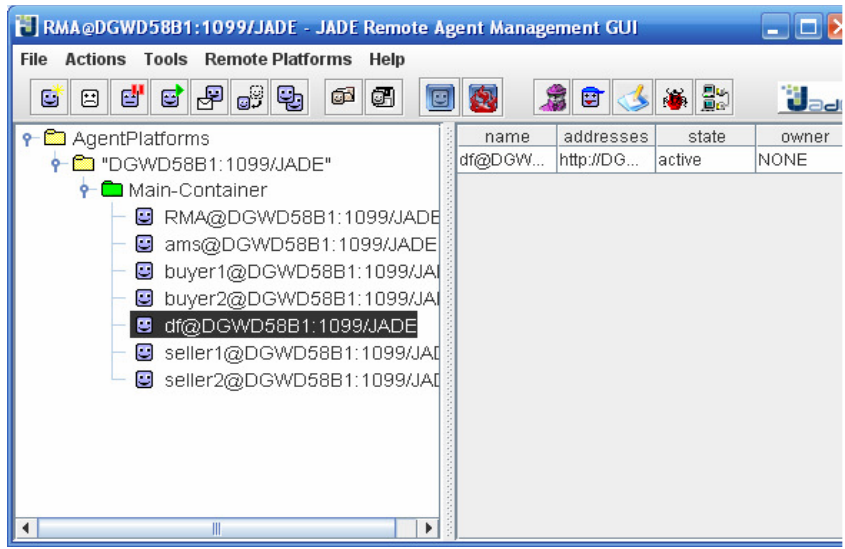

Fig.5: Jade Remote Agent Management GUI.

The above picture shows the agent management graphical user interface for JADE. The GUI is very user friendly and helps the user to handle all the operations at one console. The performance of the agents and their communications with other peer agents can be viewed by using the sniffer agent tool provided in the GUI.

A typical auction scenario including three buyers and a seller is and their communication is depicted in Fig 6. This is accomplished using Jade agents. Thus agent technology has been used to improve the performance of the present auction system. The performance of agents has been compared using the parameters of user involvement, expected result, obtained result and time taken to complete. Web sites hosting auctions with the help of Jade agents have better scalability and perform better than other auction hosting sites which use traditional non agent methodology.

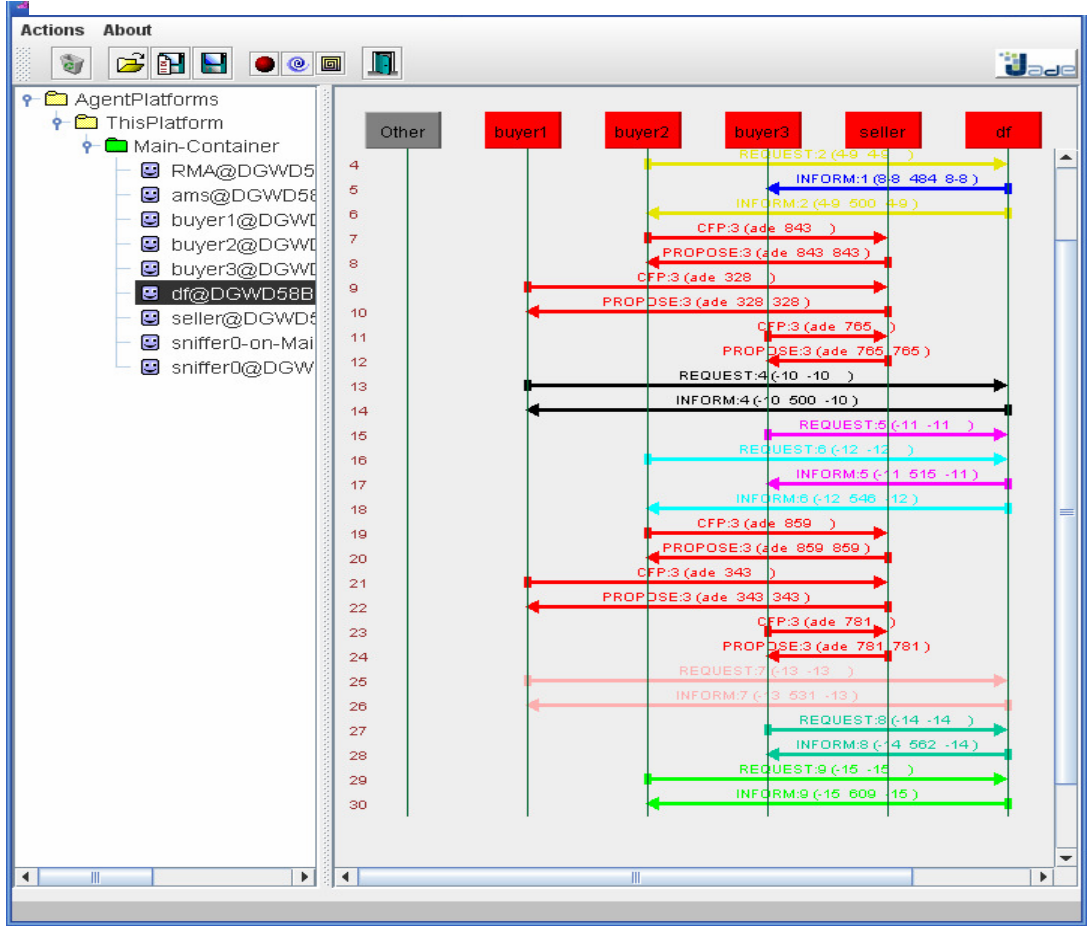

Fig. 6: Typical Auction Scenario (3 Buyers and 1 Seller)

## 4.2 Trust Management

The trust calculation is done based on the reputation of the person who is rating. Consider 'x' as the person whose weight value is needed for the trustworthiness. Let 'y' be another user who has a common partner set with 'x'. E.g. the common partner set means that suppose x and y are clients who have won auctions hosted by some common set of people. The weight value of the trust worthiness of x is calculated by comparing the ratings given by x to each and every element of the common set as compared to the ratings given to the same element by y. Several critical attributes are taken into consideration while the rating procedure is done.

(i) Weight value of x is calculated as

$$\sum_{a \in Common(x,y)} \frac{\sum_{i=1}^{N(Critical\ attributes)} fcr(i)(x,a) \cdot fcr(i)(y,a)}{\left|\sum_{i=1}^{N(Critical\ attributes)} fcr(i)(x,a)\right| + \left|\sum_{i=1}^{N(Critical\ attributes)} fcr(i)(y,a)\right|} \bigg/ N(common(x,y))$$

(ii) The Optimal price is calculated as

Initial price + $\sum_{i=1}^{n(days)} \{(0.1) \cdot (initial\ price) - (\Pi - 0.5) \cdot (priority)\}$

(iii) Time component is calculated as

$f_{((n-1)th\ auction)} - (f_{((n-1)th\ auction)} /$ time since last auction)

(iv) Trust value of x =

$e^{(weight\ value\ of\ x)(weight\ value\ of\ optimal\ price)(time\ component)(exp)}$

Here,

- $f_{cr(i)}(x,a)$ denotes the rating awarded by x to a for the critical attribute of i.
- N(common(x,y)) denotes the number of users common to x and y.
- n(days) denotes the number of days in which the auction should be complete.
- f(n-1 auction) is the feedback the person had until the last auction he hosted was complete.
- Exp is the experience value. The experience values are calculated based on the number of auctions participated and the number of auctions won.

Suppose that *x* is a client in an e auction scenario. This implies that x is a potential candidate for winning the auction and hence a potential rater. Suppose that *x* has won the auctions hosted by *a, b, c* and *d.* according to the methodology described above, another client who has won auctions which form a subset of that of x is found.

(i.e) example: Here x and y have the most common sellers which consist of the set of common$_{(x, y)}$ = {a, b, c, d} and N$_{common(x, y)}$ in this case is 4. Now to calculate the weight value of *x*, a comparison is done based on the number of critical attributes and the rating x and y gave to each element in N$_{common(x, y)}$.

|   | A | b | C | d | e |
|---|---|---|---|---|---|
| X | ☒ | ☒ | ☒ | ☒ |   |
| Y | ☒ | ☒ | ☒ | ☒ | ☒ |
| Z | ☒ |   | ☒ |   | ☒ |
| W |   | ☒ |   | ☒ | ☒ |

Table 1: Common Subset

Suppose there are 3 critical attributes $C_1$, $C_2$, $C_3$. These critical attributes vary according to the product in question. Some examples of the critical attributes are delay in delivery, the quality promised as opposed to the obtained quality, expected price as opposed to price bought etc. E.g. If the ratings of x and y towards a, b, c, d as per the critical attributes of $c_1$, $c_2$ and $c_3$ are as follows.

|   | A | | | b | | | c | | | d | | |
|---|---|---|---|---|---|---|---|---|---|---|---|---|
|   | C1 | C2 | C3 | C1 | C2 | C3 | C1 | C2 | C3 | C1 | C2 | C3 |
| X | 3.5 | 4 | 5 | 3 | 5 | 2 | 5 | 5 | 5 | 1 | 4.5 | 4 |
| y | 4 | 2 | 4 | 2.5 | 4 | 3 | 1 | 0 | 5 | 2 | 5 | 4 |

Table 2: Feedback Values

By using these values which are stored in the database for the purpose of comparison, the weight value of x is found using the first equation. Thus the numerator in the first equation turns into the following.

$$R_a = \frac{f_{(c1(x,a))} \cdot f_{(c1(y,a))} + f_{(c1(x,a))} \cdot f_{(c1(y,a))} + f_{(c1(x,a))} \cdot f_{(c1(y,a))}}{\left| f_{(c1(x,a))} + f_{(c2(x,a))} + f_{(c3(x,a))} \right| + \left| f_{(c1(y,a))} + f_{(c2(y,a))} + f_{(c3(y,a))} \right|}$$

$$R_b = \frac{f_{(c1(x,b))} \cdot f_{(c1(y,b))} + f_{(c1(x,b))} \cdot f_{(c1(y,b))} + f_{(c1(x,b))} \cdot f_{(c1(y,b))}}{\left| f_{(c1(x,b))} + f_{(c2(x,b))} + f_{(c3(x,b))} \right| + \left| f_{(c1(y,b))} + f_{(c2(y,b))} + f_{(c3(y,b))} \right|}$$

The above calculation is continued for each and every element of the set $N_{common(x, y)}$ which are b, c and d. Here $f_{(c1(x, a))}$ denotes the value 3.5 in the table 2 which means the feedback given by x to a for critical attribute of $C_1$. If the above result is considered a $R_a$ and the corresponding results for b, c and d are $R_b$, $R_c$, and $R_d$ respectively. Thus the weight value of x is calculated as follows

$$W_x = \frac{R_a + R_b + R_c + R_d}{4}$$

The critical attributes mainly revolve around the extent of customer satisfaction, the quality of the product when received as compared to that of promised quality by the vendor etc. The calculation of optimal price is required to further enhance the accuracy in the prediction of the trust values. The optimal price calculated is assigned some weight value which is based on the experience of the vendors or the hosts of the auctions. The effect of the optimal price on the trustworthiness is solely dependent on the type of tangible goods sold and the type of services provided and the type of customers who visit it.

The optimal price is calculated taking the data given by the seller. The *priority* value denotes the urgency with which the product has to be sold. The range of these values can be in the closed interval of [0, 1]. The other parameters taken from the user are the *initial price* which denotes the starting price of the auction. The $N_{(days)}$ denotes the ending date of the auction. This value calculated is used to avoid the problem of reputation squeeze.

The comparison between the trust value which is calculated and also the experience value which is obtained shows behaviour of the particular user and this can be left to the discretion of the other co users. The complexity of this proposed model is comparatively less and the processing time is also reduced since the numbers of operations made on the database are reduced. Apart from the trust model which is provided, there are lot of psychological factors that affect the trust management methodology inherently. Thus several factors which affect the performance of the site are listed below. The focus should be on avoiding distrust building factors because trust can only be built over time and multiple interactions. But a negative influence spreads in a very short span of time with very little information and negative information weighs more heavily in human judgment. Some of the factors are

- **Credibility:** Professional appearance of the website, ease of use, up-to-date information, Good interaction, photographs of the staff and several other factors contribute to credibility.

- **Security through 3rd party:** Globally recognized security must be used.

- **Added Incentives:** Trust can be built easily if the customer thinks that he has nothing to lose. So giving advices regarding what else did the other customer buy with this product, and information regarding the offers available and efficient trust model help the cause.

- **Experience:** Feedback facility should be provided to share the experiences as previous experiences which can be self or transferred play a major role in trust management.

## 5. Results

The system is successful in catering to the needs of the users by providing them the automated and less tiresome more reliable way of online shopping. The performance, scalability and efficiency of the system discussed is more compared to previous models The comparison between the performances of various auction protocols has been done in two cases. One with agents and the other without agents is considered. The plain line indicates the one with agents, the bar chart represents the expected price and the line with markers represents the one without agents. Fig.7 shows the performance when the price is taken into consideration as a key factor. We take the expected price from the optimal price formula and compare it with the final prices obtained in the system with agents and also in the one without agents. It depicts that in most cases the price at which the buyer bought the product is often less in the system which uses agent technology. This is because of the fact that there is a continuous monitoring on the auction participated by the buyer using the buyer agent and thus the negotiation strategies of the buyer are much more effective than the traditional online auctions where the buyer has to manually keep track of all the proceedings of the auction in real time.

The Fig 8 shows that the performance of the auctions with agents is much better than the one without agent technology in case of dutch protocol. The price is often marginally low in case of the auctions system which use agent technology. Unlike the english auction scenario, the difference in performance is quite evident in the dutch auction scenario.

In the dutch auction scenario the result is very evident because here we provide a range of values where a buyer can stop the negotiation process and purchase the product. But in the normal phase if the buyer is not present at the exact time at which the seller reduces the price in his favor, then some other buyer will buy the object. by using agents there is no scope of such mistakes happening.

The Fig 9 shows the variation in case of Vickrey auction scenario. The price is almost the same in both the cases with and without agents except a few minor marginal changes. But the variation in optimal price is very high and is rather unpredictable based on the graph. There are no major changes in both the systems except when the price range is at the extremeities.

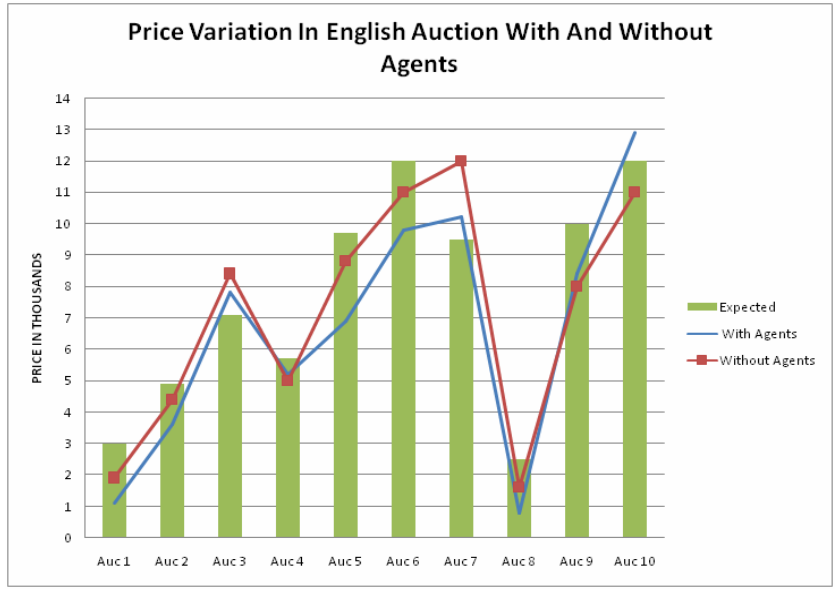

Fig.7 Price variation in English auction with and without agents.

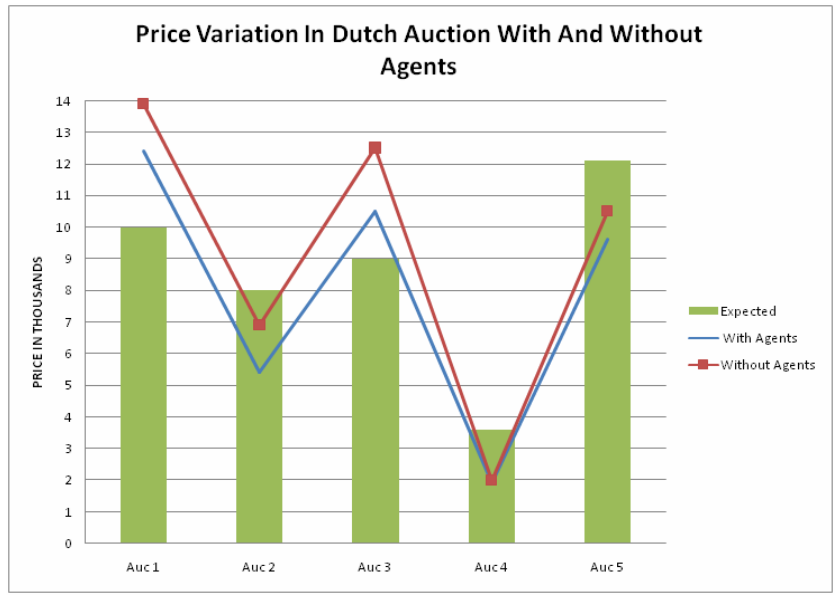

Fig.8 Price variation in dutch auction with and without agents.

Thus the performance in the case of vickery auctions taking the price factor into consideration can be highly unpredictable as they belong to the category of sealed bid auctions. Thus the most

28

effective performance in many cases has been found to be in the systems which use the agent technology. The working of the optimal price which has been proposed has been most effective in case of the English auction scenario.

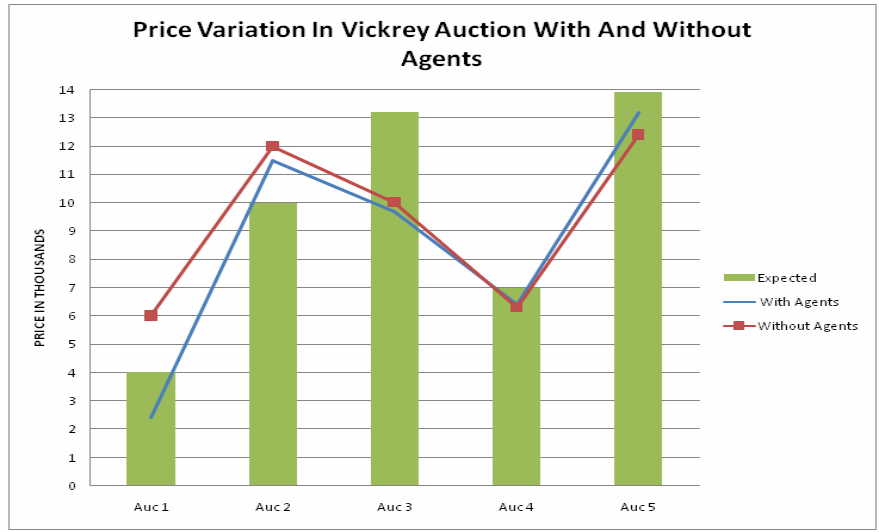

Fig.9 Price variation in vickrey auction with and wihtout agents.

## 6. Conclusion and Future work

Work is done on the task of increasing the reliability of these feedback mechanisms taking certain parameters into consideration. The impact of malicious feedbacks is reduced. There are many other real time factors which are necessary to be taken into consideration to depict the dynamic nature of the present world scenario of e-auctions. To generalize the agents to participate in other types of auctions such as stock exchanges, which are asynchronous double auctions would be a good future prospective. Another future enhancement will be to add additional bidding strategies. Detailed performance statistics will be collected to determine which strategies perform better under which types of auctions. Comparison of these strategies with the human strategies is also an option.

**Authors**

E.Sathiyamoorthy is an Assistant professor (SG) in school of Computing sciences at VIT University, Tamilnadu, India. He received his MCA degree from University of Madras. He is pursuing his PhD at VIT University Vellore. His Research interests include E-Business, Agent Technologies and Web Services. He attended many conferences and workshops and published papers.

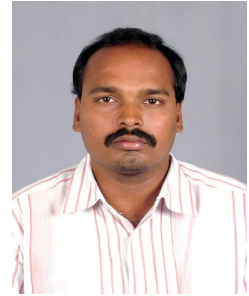

Dr.N.Ch.S.N.Iyengar is a Senior Professor at the School Of Computing Sciences at VIT University, Vellore, Tamilnadu  India. He received M.Sc (Applied Mathematics) & PhD from Regional Engineering College Warangal (Presently known as NIT Warangal).Kakatiya University, Andhra Pradesh, India, & M.E. (Computer Science and Engineering) from Anna University, Chennai, India. His research interests include Fluid Dynamics (Porus Media), Agent based E-Business Applications, Data Privacy, Image Cryptography, Information security, Mobile Commerce and cryptography. He has authored several textbooks and had  research Publications  in National , International Journals & Conferences. He is also   Editorial Board member for many National and International Journals. He chaired many International conferences' and delivered invited , technical lectures  along with keynote addresses beside being . International programme committee  member.

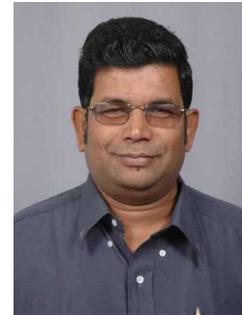

Dr.V. Ramachandran, presently Vice–Chancellor of Anna University, Trichy (Tamil Nadu). He served as Professor in the Department of Computer Science at the College of Engineering, Anna University, India. He received his ME and PhD Degrees from Anna University, India in 1982 and 1991, with specialisation in Power Systems. He served as visiting professor in several national and international institutes. He has authored several research publications. He chaired many International conferences' and delivered invited  technical lectures  along with keynote addresses .His research interests include power systems analysis in distributed environment, networks and web technologies. He is serving as an editorial member of many international and national journals.

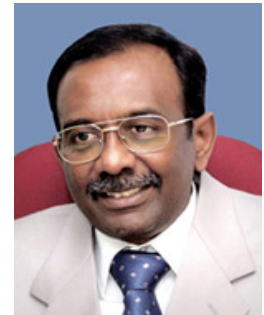